\documentclass[english,prl,two column, superscriptaddress]{revtex4-1}
\usepackage[T1]{fontenc}
\usepackage[latin9]{inputenc}
\usepackage{amsmath}
\usepackage{graphicx}
\usepackage{amssymb}

\makeatletter

\@ifundefined{textcolor}{}
{%
 \definecolor{BLACK}{gray}{0}
 \definecolor{WHITE}{gray}{1}
 \definecolor{RED}{rgb}{1,0,0}
 \definecolor{GREEN}{rgb}{0,1,0}
 \definecolor{BLUE}{rgb}{0,0,1}
 \definecolor{CYAN}{cmyk}{1,0,0,0}
 \definecolor{MAGENTA}{cmyk}{0,1,0,0}
 \definecolor{YELLOW}{cmyk}{0,0,1,0}
 }

\newcommand{\carbon}{$^{13}$C}
\newcommand{\siliconspin}{$^{29}$Si}

\newcommand{\tone}{$T_1$}

\newcommand{\proton}{$^{1}$H}
\newcommand{\mw}{$\mu$W}
\makeatother

\usepackage{babel}
\usepackage{epsfig}

\usepackage[bookmarks]{hyperref}

\begin{document}

\title{Radical-free dynamic nuclear polarization using electronic defects in silicon}

\author{M.~C.~Cassidy}
\affiliation{School of Engineering and Applied Sciences, Harvard University, Cambridge, Massachusetts 02138, USA }
\author{C.~Ramanathan}
\affiliation{Department of Physics and Astronomy, Dartmouth
College, Hanover, NH 03755, USA}
\author{D.~G.~Cory}
\affiliation{Department of Chemistry and Institute for Quantum
Computing, University of Waterloo, Waterloo ON N2L 3G1, Canada}
\affiliation{Perimeter Institute for Theoretical Physics, Waterloo, ON
N2L 2Y5, Canada}
\author{J.~W.~Ager}
\affiliation{Materials Sciences Division, Lawrence Berkeley National Laboratory, Berkeley, California 94720, USA}
\author{C.~M.~Marcus}
\affiliation{Department of Physics, Harvard University, Cambridge, Massachusetts 02138, USA}
\affiliation{Center for Quantum Devices, Niels Bohr Institute, University of Copenhagen, DK-2100, Denmark }

\begin{abstract}
Direct dynamic nuclear polarization of \proton~nuclei in frozen water and water-ethanol mixtures is demonstrated using silicon nanoparticles as the polarizing agent. Electron spins at dangling-bond sites near the silicon surface are identified as the source of the nuclear hyperpolarization.  This novel polarization method open new avenues for the fabrication of surface engineered nanostructures to create high nuclear-spin polarized solutions without introducing contaminating radicals, and for the study of molecules adsorbed onto surfaces. 

\end{abstract}
\maketitle

Dynamic nuclear polarization (DNP) has recently emerged as a powerful technique for improving the sensitivity of nuclear magnetic resonance (NMR) experiments, leading to new insights in materials characterization \cite{Lelli:vo, Reynhardt:1998uz, Dementyev:2008up, Itahashi:2009vw}, complex biomaterials \cite{Hall:1997tu, vanderWel:2006vya, Bajaj:2009uf, Armstrong:2011gp}, and diagnostic medicine using magnetic resonance imaging (MRI) \cite{Golman:2000wp, Day:2007tf, Gallagher:2008ui, Wilson:2009we}. By applying microwave (\mw) irradiation at or near the electron Larmor frequency, a sizable nuclear spin polarization can be generated via the transfer of electron spin polarization to nearby nuclear spins. The effect can be enhanced by operating at low temperatures or high magnetic fields, where the equilibrium electron polarization can approach unity. The electrons used in the polarization process may originate from free radicals or paramagnetic metal ions introduced into the system, or be intrinsic to the material being polarized. Both the rate and saturation magnitude of nuclear polarization depend strongly on the concentration of unpaired electrons. This is because spin transport via nuclear spin diffusion is slow \cite{Abragam:1978wp,Bloembergen:1949uw} and competing relaxation processes means that a high radical concentration is required to achieve large polarizations on experimentally relevant time scales. 

Most biological applications of DNP involve systems that have no intrinsic unpaired electrons for polarization, and so a significant effort has been undertaken to develop a range of free radicals \cite{AL,Hu:2007wa, Song} that can be dissolved in a solvent (such as water, ethanol, or glycerol) along with the biomolecular substrate to be polarized before freezing for DNP. However, the presence of these radicals in the polarized substrate after polarization reduces the spin lattice relaxation time (\tone) of the polarized substrate \cite{Mieville:2010fa, Kurdzesau:2008dg}, interferes with the chemical environment of molecules under study \cite{BDArmstrong:2007va} and, for biological experiments, poses significant toxicity issues.  To reduce contamination of the sample by residual radicals, filtering \cite{Filtering} and radical scavenging \cite{Mieville:2010fa} undertaken post-disolution have been attempted with limited success. The radicals may be bound into a porous organic matrix \cite{Dorn:1991tg,McCarney:2007ux}, though  filtration is still necessary. Alternatively, the source of polarization may be physically separated from the material to be polarized. It has been shown that the large nuclear spin polarizations present in optically pumped noble gases and semiconductors can be transferred to a variety of materials using a Hartmann-Hahn cross-polarization sequence \cite{Tycko, Raftery:1993vr, Goehring:2003tf, Goto:2008we} or, in the case of gases, via a direct nuclear dipole-dipole interaction \cite{Bowers:1993vp}.  Surface defects in porous carbon chars have also been shown to polarize \proton~nuclei in solution at room temperature via a contact hyperfine interaction \cite{odintsov}.   

In this Letter,  we demonstrate that naturally occurring paramagnetic defects at the surface of silicon nanoparticles (SiNP) can directly polarize \proton~nuclei in a bulk frozen matrix near the surface of the particle as well as the \siliconspin~nuclei in the particles. Two distinct nuclear spin baths are observed, one consisting of nuclei polarized directly by  surface electrons, and a second consisting of nuclei polarized indirectly by nuclear spin diffusion. By comparing natural abundance and isotopically enriched samples, we find that \proton~polarization is strongly affected by the isotopic concentration of \siliconspin~nuclei inside the particles. This technique allows uncontaminated hyperpolarized solutions to be generated at room temperature, benefitting applications from polarized targets to molecular imaging. 

\begin{figure}[ht]
\centering{}\includegraphics{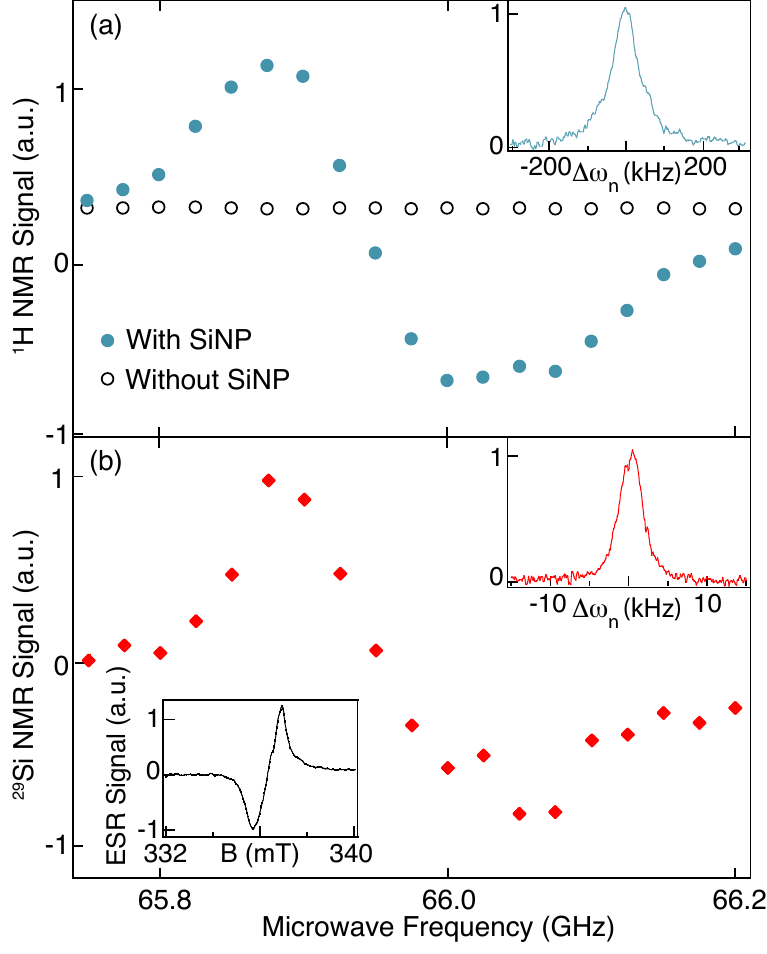}
\caption{Microwave frequency dependence of the (a) \proton~NMR signal (closed circles) and (b) \siliconspin~ NMR signal from a sample of natural isotopic abundance SiNP suspended in frozen water, measured at 4.2~K. No change in the \proton~NMR signal was observed for the same sample without the SiNP (open circles), and no \proton~signal was observed from a dry sample of SiNP. Upper insets in (a) and (b): \proton~and \siliconspin~spectra for \mw~irradiation frequency of 65.875~GHz. Lower inset in (b): Room temperature ESR signal of natural abundance SiNP.}
\end{figure}

 \begin{figure}[ht]
\includegraphics{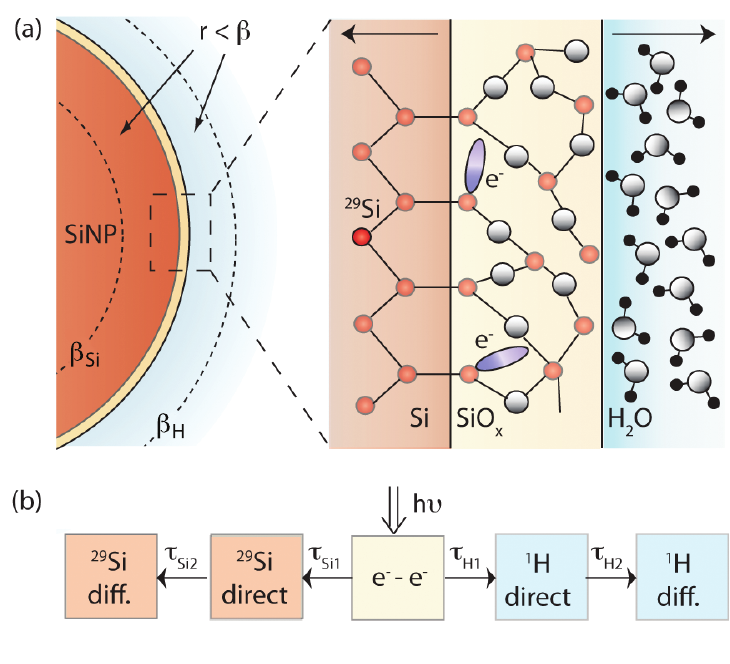}
\caption{(a) Schematic model showing the location of DNP-active electron spins at and near the silicon-silicon oxide interface. Arrows indicate the directions of nuclear polarization flow away from the surface of the particle. (b) Thermodynamic model (see text) for the polarization process of both the \proton~and \siliconspin~ nuclear spin baths. \siliconspin~and \proton~nuclei near the defect electrons are polarized directly through a dipolar interaction on time scales $\tau_{Si1}$ and $\tau_{H1}$. Nuclei far away from the surface are then polarized by nuclear spin diffusion on time scales $\tau_{Si2}$ and $\tau_{H2}$ that are dependent on the spatial separation of the nuclear spins.}
\end{figure}

Samples consisted of natural abundance (\siliconspin~=~$4.8\%$) polycrystalline powder (Alfa Aesar, $d = 3~\mu \rm{m}$),  and isotopically enriched (\siliconspin~=~$91.4\%)$ \cite{Ager} silicon particles of diameter $d = 200~\rm{nm}$ made by ball milling bulk silicon and then size-separated by centrifugal sedimentation \cite{Aptekar:wx}. Particle sizes were confirmed by scanning electron microscopy (SEM). Suspensions of these particles in water (1:1 Si:$\rm{H_{2}0}$ by weight), and ethanol and water (10:9:1 Si:EtOH:$\rm{H_{2}0}$ by weight) were prepared by sonication and then degassed with a freeze-pump-thaw cycle before being loaded into the cryostat for measurement.

DNP experiments were performed at 4.2~K using a continuous flow cryostat in an applied field, $B$, of 2.35~T ($\omega_{n}^{\rm{29Si}}$ = 20 MHz, $\omega_{n}^{\rm{1H}}$ = 100 MHz). The sample was housed in a glass capillary tube in direct contact with the flowing helium vapor.  A Bruker NMR spectrometer with a home-built probe and solenoidal coil \cite{Cho:2007ua} were used for NMR detection. Microwave irradiation was provided through a 90~mW Gunn diode source (Millitech), and the \mw s were coupled from room temperature to the sample via a mm waveguide and horn antenna. The nuclear polarization was measured with a saturation recovery sequence (\siliconspin)  $\left(\pi/2\right)^{16} - t - \left(\pi/2\right)_{X}$  or saturation recovery solid echo sequence (\proton)  $\left(\pi/2\right)^{16}  - t - \left(\pi/2\right)_{X} - t_{d} - \left(\pi/2\right)_{Y}$ due to the rapid decay of the \proton~signal, corresponding to a $\sim 100$ kHz \proton~dipolar linewidth.

Figure 1(a) shows the \mw~frequency dependence of the \proton~NMR signal for solid H$_2$O samples, with and without suspended SiNP (natural isotopic abundance), for a \mw~irradiation time of 120~s. For the sample with SiNP, the \proton~nuclear polarization depended strongly on the applied \mw~frequency, while no change in \proton~polarization was seen for the sample without SiNP. A similar dependence on \mw~frequency of the \siliconspin~NMR signal for the same sample of SiNP suspended in frozen water (Fig. 1(b)) after 600~s of \mw~irradiation. No \proton~signal was observed from a sample of dry SiNP. Representative \proton~and \siliconspin~spectra at a \mw~frequency of 65.875 GHz are shown in the insets \cite{spectrum}. We note that no change in the room temperature \proton~spectrum was seen in a similarly prepared sample after repeated measurements over a period of a week. This suggests that chemical reactions at the silicon/water interface leading to Si-H bond formation is insignificant, and that the \proton~NMR signal results from \proton~nuclei outside of the particle rather than absorbed into the oxide or particle itself.

Both \proton~and \siliconspin~polarizations show inversions at the same \mw~frequency, $\omega_{0} = 65.95$~GHz, indicating that polarization of both nuclear species originates from electrons with the same g-factor. Room temperature X-band electron spin resonance (ESR)  (lower inset of Fig.~1~(b)) of the particles shows a single broad peak at $B = 335.9$~mT and frequency 9.444 GHz, corresponding to g = 2.006.  This g-factor is consistent with the orientation-averaged signal from unpaired electrons at $P_{b}$ defects, which are known to occur at the silicon-silicon dioxide interface \cite{Caplan:1976wk}. To confirm the role of these defects in the polarization process we prepared a similar sample where the surface oxide was removed by a hydrofluoric acid etch in an inert atmosphere and suspended in glycerol. This sample showed no ESR signal and no \proton~or \siliconspin~DNP enhancement under similar conditions. Additionally, the sign of polarization enhancement is the same for both nuclear species for frequencies either side of $\omega_{0}$, despite the gyromagnetic ratios of the \proton~ and \siliconspin~ nuclei being opposite in sign. From this we can determine that the \proton~nuclei are being polarized through a direct interaction with the electron dipolar spin bath, rather than via a dipolar interaction with polarized \siliconspin~nuclei close to the surface.  The direct interaction scales $\propto \gamma_{n}^2$ \cite{Khutsishvili:1955vd}, independent of the sign of $\gamma_{n}$, the nuclear gyromagnetic ratio.  If instead the \proton~polarization was mediated by a dipolar interaction with the polarized \siliconspin~nuclei, opposite nuclear polarizations would be generated in the two nuclear species for the same polarizing frequency.

A schematic model of the system is shown in Fig.~2. The anisotropic g factor of the $P_{b}$ defect  ($\rm{g} = 2.002 - 2.010$) \cite{Caplan:1976wk}, together with the high density of unpaired electrons at the sample surface ($\sim10^{13} \rm{cm}^{-2}$) \cite{MYL} yields a broad spectrum of electron-electron dipolar interactions with energies at or near the Lamor frequency of each nuclear species. This results in a broad \mw~frequency response of both \proton~and \siliconspin~nuclei (Fig. 1), consistent with previous DNP studies of silicon microparticles \cite{Dementyev:2008up}.

\begin{figure}[tbp]
\includegraphics{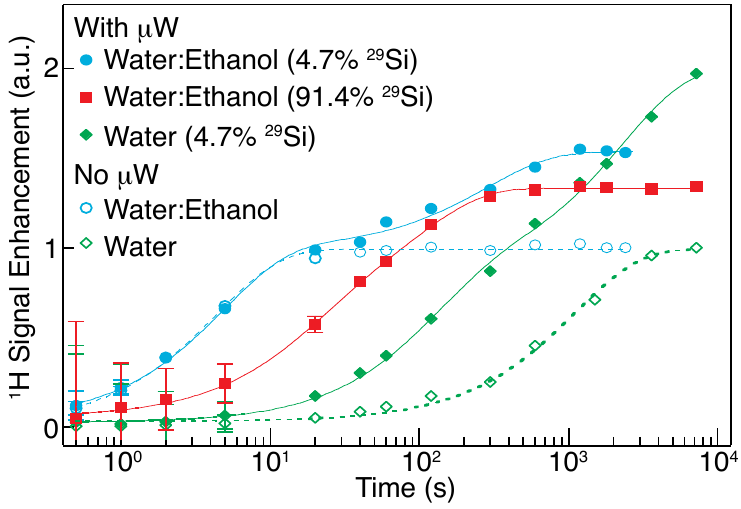}
\caption{Time evolution of the \proton~nuclear polarization with (solid shapes) and without (open shapes) applied \mw~irradiation for frozen solutions of $\rm{SiNP:H_20}$ (4.7\% \siliconspin) and $\rm{SiNP:EtOH:H_20}$ (4.7\% \siliconspin, 91.4\% \siliconspin). The enhancements are scaled relative to the 4.2K equilibrium polarization. Fits are exponential (dashed lines) or biexponential (solid lines), see text. Error bars are shown when greater than the data symbol.}
\end{figure}

The time evolution of the \proton~polarization of various frozen suspensions of natural abundance and \siliconspin-enriched SiNP in frozen water and water-ethanol mixtures at \mw~frequency of 65.875~GHz is shown in Fig.~3. All data are normalized to the \proton~equilibrium polarization ($P_{B}$) at 4.2~K without \mw~irradiation. The enhancement curves with \mw~irradiation display biexponential behavior, well fit by the form  $P=P_{\infty}\left(1-(Ae^{-t_{\rm pol}/\tau_{1}}+Be^{-t_{\rm pol}/\tau_{2}})\right)$ (solid lines).  This is in contrast to the time evolution of the \proton~NMR signals without \mw~irradiation, which are better described by a single exponential, $P=P_{B}\left(1-e^{-t_{\rm pol}/T_{1}}\right)$ (dashed lines). Best fit values for these characteristic times are given in Table 1. 

Note in Fig.~3 that the long-time induced polarization in the water-ethanol sample is suppressed relative to the water sample. We attribute this to fast nuclear relaxation caused by the rotating methyl group in ethanol, which creates an oscillating magnetic field with a characteristic time constant of a few seconds even at cryogenic temperatures  (methyl rotamer effect \cite{Clough:1982tn}). Increasing the \siliconspin~isotopic concentration from $4.7 \%$ to $91.4\%$ and reducing the particle size from 3 $\mu$m to 200 nm reduces the polarization rates and the total \proton~polarization enhancement by a factor of two, from 1.5 to 1.25 times the Boltzmann polarization at 4.2~K. This is particularly notable given the higher number of polarization sites arising from surface area to volume ratio that is 15 times larger in these samples. Despite this significant increase in the availability of directly polarization sites, the higher concentration of \siliconspin~nuclear spins in the NP are able to absorb a greater total angular momentum transfer from the surface electrons compared to the natural abundance samples. This results in a slower direct polarization rate of the \proton~nuclear spins close to the surface. However, the relaxation due to the methyl rotamer effect and other electronic defects within the frozen matrix remains the same as for the natural abundance samples, and so competes more strongly with polarization by spin diffusion and hence causes the \proton~polarization to saturate earlier and at a lower absolute value.

\begin{table}
\centering
\begin{tabular}{ l l | c | c  | c }
\hline
\hline
Sample && $T_{1} (\rm{s})$&$\tau_{\rm{H}1} (\rm{s})$& $\tau_{\rm{H}2} (\rm{s})$  \\ 
\hline
Si:$\rm{H_{2}0}$& (4.7\% \siliconspin)&$1113\pm96$ & $120\pm14$ & $2304\pm307$ \\
Si:EtOH:$\rm{H_{2}0}$ &(4.7\% \siliconspin)&$4.3\pm0.4$ & $4.6\pm0.8$ & $217\pm58$\\
Si:EtOH:$\rm{H_{2}0}$ &(91.4\% \siliconspin)&  &$16\pm3$ & $97\pm17$ \\
\hline
\hline
\end{tabular}
\caption{\proton~polarization times fitting to experimental data in Fig. 3. $T_{1}$ is the characteristic time taken for the \proton~nuclei to return to the equilibrium polarization without \mw~irradiation. $\tau_{H1}$ and $\tau_{H2}$ are fast and slow polarization times which form two additive components of the polarization process corresponding to polarization dominated by direct DNP and spin diffusion respectively.}
\label{table:nonlin-H}
\end{table}

It is possible to estimate the size of the region of direct polarization for each nuclear species. Upon \mw~irradiation of the electron spin bath, the polarization of nuclei within a radius $r < \beta$ is through a direct dipolar interaction with a pair of electrons at a rate $\tau_{1}^{-1}=\Gamma/r^6$, where $\Gamma =  KCG(\omega_{0}-\omega^*)$, 
$K =  \pi\gamma_{e} B_1^2 \tau/2$ is the saturation strength of the microwave field,
\begin{align} C = & \frac{3}{10}\frac{\hbar^{2}\gamma_{e}^{2}}{B^{2}}  \frac{B^{2}\gamma_{n}^{2}T_{2e}}{1+ B^{2}\gamma_{n}^{2}T_{2e}^2}, \end{align} 
is a constant describing the nature of the electron-nuclear dipolar interaction, and 
$G(\omega-\omega^*)$ is the electron lineshape function \cite{Khutsishvili:1955vd}. Here $B_1$ is the strength of the \mw~magnetic field, $\tau$ is the correlation time of the electron spin magnetization orientated along the direction of the static magnetic field ($1/\tau=1/T_{1e}+1/T_{2e}$), $T_{1e}$ is the electron spin lattice relaxation time and $T_{2e}$ the transverse electron spin relaxation time. The distance $\beta=(C/D)^{1/4}$ (D is the spin-diffusion constant), characterizes the cross-over radius between direct and diffusion-mediated hyperpolarization \cite{Khutsishvili:1955vd} .

For nuclei located outside this radius (r $> \beta$), polarization occurs predominantly via nuclear spin diffusion with a rate $\tau_{2}^{-1} = \frac{\partial P}{\partial t} = D\nabla^{2}P$ \cite{Bloembergen:1949uw}. $D=Wa^{2} \sim a^{2}/(50\,T_{2n})$ is the spin-diffusion constant, with $a$ the average separation between nearest-neighbor nuclei, $W$ the probability of a flip-flop transition between nuclei due to dipole-dipole interaction, and $T_{2n}$ the transverse nuclear spin relaxation time \cite{Khutsishvili:1955vd}. For the present system, we estimate $\beta_{\rm{H}} \simeq 3~\rm{nm}$ and $\beta_{\rm{Si}} \simeq 4~\rm{nm}$, given $T_{1e} \simeq 30 \mu \rm{s}$ and $T_{2e} \simeq 1~\rm{ns}$ \cite{Dementyev:2008up}, $a_{\rm{H(Si)}} \approx 1.3~(4.1)$ {\AA}, and  $T_{2n\rm{H(Si)}} = 0.01~(5.6)~\rm{ms}$. Transmission electron micrograph studies of the SiNP show an oxide thickness of a few nm, a distance small enough for this direct polarization process to take place. We note that $\beta$ is not a hard cutoff of the distance for direct polarization to take place, rather it gives a transition point between polarization dominated by direct DNP and spin diffusion. 

Although the total enhancement (ranging from $\sim1.5-3$) over all \proton~ spins in the sample is relatively low compared to enhancements in concentrated radical solutions \cite{ArdenkjaerLarsen:2008tb}, the enhancement near the surface of the particles is significant. Numerical modeling of the polarization as a function of distance from the particle surface shows a characteristic length of order 10~nm. This is consistent with studies of polarization as a function of radical concentration \cite{Kurdzesau:2008dg}. Estimating the enhanced polarization of \proton~nuclei within 10~nm of the surface of the SiNP (corresponding to $\sim1.2\%$ of all \proton~nuclei in the sample) based on the model above, we find a factor of $\sim 50$ above equilibrium polarization at 4.2~K for the natural abundance particles in the ethanol solution, corresponding to an enhancement of $\sim~3000$ times the room temperature equilibrium polarization. This enhancement exceeds the reported enhancements of functional groups on silica surfaces using artificial free radicals and much higher power microwave sources, estimated at $\sim 30$ times the equilibrium polarization at 77~K for molecules within 1~nm of the surface \cite{Lesage:2010cx}. 
  
This effect may be optimized for enhanced solution polarization or studies of surface functionalization by reducing the \siliconspin~concentration, increasing the surface area of the SiNP, using porous silicon, or engineering thinner surface oxides with higher defect densities. Additionally we note that this effect may be revealed in other solid state systems with a lower nuclear spin concentration or lower gyromagnetic ratio. In particular, \carbon~DNP has been demonstrated in a variety of diamond materials \cite{Reynhardt:1998uz} ($1\%$ \carbon~natural abundance) and nanodiamond and has surface active electronic defects \cite{Shames} that may be suitable as polarizing agents by mechanisms similar to those described in this paper.  

We acknowledge support from the National Science Foundation under NSF-0702295, the BISH Program (CBET-0933015), the Harvard NSF Nanoscale Science and Engineering Center, and the Canada Excellence Research Chairs Program. Fabrication used the Harvard Center for Nanoscale Systems (CNS), an NSF National Nanotechnology Infrastructure Network (NNIN) site (ECS 0335765). Work at the LBNL (\siliconspin~synthesis) was supported by the Director, Office of Science, Office of Basic Energy Sciences, Materials Sciences and Engineering Division of the US Department of Energy (DE-AC02-05CH11231).


\begin{thebibliography}{10}

\bibitem{Lelli:vo} 
\newblock M. Lelli \it{et al.}\rm, J. Am. Chem. Soc., \bf133\rm, 2104-2107 (2011).
\bibitem{Reynhardt:1998uz}
\newblock E. Reynhardt, and G. J. High, J. Chem. Phys. \bf 109 \rm, 4090-4099 (1998).
\bibitem{Dementyev:2008up}
\newblock A. Dementyev, D. Cory, and C. Ramanathan, Phys. Rev. Lett. \bf 100 \rm, 127601 (2008).
\bibitem{Itahashi:2009vw}
\newblock H. Hayashi \it{et al.}\rm, Phys. Rev. B. \bf 80 \rm, 045201 (2009).
\bibitem{Bajaj:2009uf}
\newblock V.S. Bajaj \it{et al.}\rm, Proc. Nat. Acad. Sci. \bf106\rm, 9244 - 9249 (2009).  
\bibitem{vanderWel:2006vya} 
\newblock P. van der Wel \it{et al.}\rm, J. Am. Chem. Soc. \bf128 \rm, 10840-10846 (2006).
\bibitem{Armstrong:2011gp} 
\newblock B.D. Armstrong \it{et al.}\rm, J. Am. Chem. Soc. \bf133 \rm, 5987-5995 (2011).
\bibitem{Hall:1997tu}
\newblock D. Hall \it{et al.}\rm, Science \bf276 \rm, 930-932 (1997).
\bibitem{Day:2007tf} 
\newblock S. Day \it{et al.}\rm, Nat. Med. \bf13\rm, 1382-1387 (2007).
\bibitem{Golman:2000wp} 
\newblock K. Golman \it{et al.}\rm, Cancer Res. \bf66\rm, 10855-10860 (2006).
\bibitem{Gallagher:2008ui} 
\newblock F. Gallagher \it{et al.}\rm, Nature, \bf453\rm, 940-943 (2008).
\bibitem{Wilson:2009we}
\newblock D. Wilson \it{et al.}\rm, Proc. Nat. Acad. Sci. \bf106\rm, 5503-5507 (2009).
\bibitem{Abragam:1978wp}
\newblock A. Abragam, \emph{Principles of Nuclear Magnetism} (Oxford University Press, 1983), p. 144.
\bibitem{Bloembergen:1949uw}
\newblock N. Bloembergen, Physica \bf15\rm, 178-219 (1949).   
\bibitem{Hu:2007wa}
\newblock K. Hu, V. Bajaj, M. Rosay, and R.G. Griffin, J. Chem. Phys. \bf126\rm, 044512 (2007).
\bibitem{Song}
\newblock C. Song \it{et al.}\rm, J. Am. Chem. Soc. \bf128\rm, 11385-11390 (2006).
\bibitem{AL}
\newblock J. Ardenkjaer-Larsen \it{et al.}\rm, Proc. Nat. Acad. Sci. \bf100\rm, 10158 - 10163 (2003).
\bibitem{Mieville:2010fa}
\newblock P. Mieville \it{et al.}\rm, Angew. Chem. \bf122\rm, 6318-6321 (2010).
\bibitem{Kurdzesau:2008dg}
\newblock F. Kurdzesau \it{et al.}\rm, J. Phys. D: Appl. Phys. \bf41\rm, 155506 (2008).
\bibitem{BDArmstrong:2007va}
\newblock E.R. McCarney, B.D. Armstrong, M.D. Lingwood, and S. Han, Proc. Nat. Acad. Sci. \bf104\rm, 1754-1759 (2007).
\bibitem{Filtering}
\newblock A. M. Leach, P. Miller, E. Telfeyan, and D.B. Whitt, (General Electric), USA US2009263325A1, (2009).
\bibitem{Dorn:1991tg}
\newblock H. C. Dorn, T. E. Glass, R. Gitti, and K. H. Tsai, Appl. Magn. Res. \bf2\rm, 9-27 (1991).
\bibitem{McCarney:2007ux}
\newblock E. McCarney, and S. Han, J. Magn. Res. \bf190\rm, 307-315 (2008).
\bibitem{Raftery:1993vr} 
\newblock D. Raftery \it{et al.}\rm, J. Phys. Chem. \bf97\rm, 1649-1655 (1993).
\bibitem{Goehring:2003tf} 
\newblock L. Goehring, and C.J. Michal, Chem. Phys. \bf119\rm, 10325-10328 (2003). 
\bibitem{Goto:2008we}
\newblock A. Goto, T. Shimizu, K. Hashi, and S. Ohki, Appl. Phys. A \bf93\rm, 533-536 (2008).
\bibitem{Tycko} 
\newblock R. Tycko, Solid State Nucl. Magn. Reson. \bf11\rm, 1 (1998).
\bibitem{Bowers:1993vp} 
\newblock C. Bowers \it{et al.}\rm, J. Mag. Res. \bf135\rm, 435-443 (1998).
\bibitem{odintsov}
\newblock B.M. Odintsov \it{et al.}\rm, J. Phys. Chem. \bf97\rm, 1649-1655 (1993).
\bibitem{Ager}
\newblock J.W. Ager \it{et al.}\rm, J. Electrochem. Soc. \bf152\rm, 448-451 (2005).
\bibitem{Aptekar:wx}
\newblock J. Aptekar \it{et al.}\rm, ACS Nano. \bf3\rm, 4003-4008 (2009).
\bibitem{Cho:2007ua}
\newblock H. Cho \it{et al.}\rm, J. Magn. Res. \bf187\rm, 242-250 (2007).
\bibitem{spectrum}
\newblock The narrow ($\sim500$ Hz) peak in the otherwise broad ($\sim2$ kHz) \siliconspin~spectrum emerges only for times > 600~s. This peak corresponds to \siliconspin~nuclei located in crystalline sections of the SiNP that are polarized by nuclear spin diffusion from \siliconspin~nuclei (that make up the broad line) located near the polarizing electrons.
\bibitem{Caplan:1976wk}
\newblock P.J. Caplan, J. N. Helbert, B.E. Wagner, and E.H. Poindexter, Surf. Sci. \bf54\rm, 33-42 (1976). 
\bibitem{MYL}
\newblock The surface defect density was estimated by comparing the relative size of the ESR spectrum to that from a bulk Si:P sample co-located in the cavity. 
\bibitem{Clough:1982tn}
\newblock S. Clough \it{et al.}\rm, J. Phys. C: Solid State Phys. \bf15\rm, 2495-2508 (1982).
\bibitem{Khutsishvili:1955vd}
\newblock G. Khutsishvili, Usp. Fiz. Nauk. \bf87\rm, 211-254 (1965).
\bibitem{ArdenkjaerLarsen:2008tb}
\newblock J. Ardenkjaer-Larsen, S. Macholl, and H. Johannesson, Appl. Magn. Res. \bf34\rm, 509-522 (2008).
\bibitem{Lesage:2010cx}
\newblock A. Lesage \it{et al.}\rm, J. Am. Chem. Soc. \bf132\rm, 15459-15461 (2010).
\bibitem{Shames}
\newblock A.I. Shames \it{et al.}\rm, J. Phys. Chem. Solids \bf63\rm, 1993-2001 (2002). 

\end{thebibliography}
\end{document}